\documentclass[twocolumn,amsmath,amssymb,floats,superscriptaddress,nofootinbib,10pt]{revtex4-1}

\pdfoutput=1

\AtBeginDocument{%
    \newwrite\bibnotes
    \def\bibnotesext{Notes.bib}
    \immediate\openout\bibnotes=\jobname\bibnotesext
    \immediate\write\bibnotes{@CONTROL{REVTEX41Control}}
    \immediate\write\bibnotes{@CONTROL{%
    apsrev41Control,author="08",editor="1",pages="1",title="0",year="1"}}
     \if@filesw
     \immediate\write\@auxout{\string\citation{apsrev41Control}}%
    \fi
}%

\usepackage{graphicx}
\usepackage{dcolumn}
\usepackage{bm}
\usepackage{revsymb}
\usepackage[usenames]{color}
\usepackage{color}
\usepackage{amsmath}
\usepackage[usenames]{color}
\usepackage{amsfonts}
\usepackage{graphicx}
\usepackage{dcolumn}
\usepackage{bm}
\usepackage{revsymb}
\usepackage[usenames]{color}
\usepackage{graphicx}

\usepackage{color}
\usepackage{amsmath}
\usepackage{amssymb,bbm}
\usepackage{subfig}

\usepackage{xcolor}

\newcommand{\bea}{\begin{eqnarray}}
\newcommand{\eea}{\end{eqnarray}}

\newcommand{\odd} {{\text{o}}}
\newcommand{\even} {{\text{e}}}

\newcommand{\sphere} {{\text{S}}}
\newcommand{\obl} {{\text{O}}}
\newcommand{\prol} {{\text{P}}}

\newcommand{\sh} {{\text{s}}}

\definecolor{nblue}{RGB}{28,130,185}

\definecolor{cgreen}{RGB}{76,153,0}

\definecolor{myorange}{RGB}{245,156,74}

\usepackage{booktabs}

\usepackage{tikz}

\usepackage{hyperref}
\hypersetup{
  colorlinks=true,
  citecolor=magenta,
  urlcolor=-myorange
}

\usepackage{xcolor}
\definecolor{ogreen} {RGB}{71,191,145}

\usepackage{hyperref}
\hypersetup{
  colorlinks=true,
  citecolor=magenta,
  urlcolor=-myorange
}

\definecolor{oblue} {RGB}{50,50,250}

\newcommand{\quotes}[1]{``#1''}

\begin{document}

\title{Sorting prolate and oblate spheroids with a diatomic gas in a magnetic field}

\author{Ruben Lier}
\email{r.lier@uva.nl}
\affiliation{Institute for Theoretical Physics, University of Amsterdam, 1090 GL Amsterdam, The Netherlands}
\affiliation{Dutch Institute for Emergent Phenomena (DIEP), University of Amsterdam, 1090 GL Amsterdam, The Netherlands}
\affiliation{Institute for Advanced Study, University of Amsterdam, Oude Turfmarkt 147, 1012 GC Amsterdam, The Netherlands}

\begin{abstract}
For a gas of diatomic particles with a nonzero magnetic moment, the 
Senftleben-Beenakker effect implies that transport can be affected by a 
magnetic field even when the particles are charge neutral. In such a gas, 
two independent anisotropic odd viscosities are allowed by symmetry and 
can have opposite signs. \textcolor{black}{We study the resulting odd-viscous Stokes flow 
around spheroids in the regime where the odd viscosities are small 
compared to the shear viscosity. Using a perturbative expansion around 
the spherical solution and the Lorentz reciprocal theorem, we compute 
the lift forces on oblate and prolate spheroids for general values of the 
two odd viscosity coefficients. We show that the resulting Hall angles 
can differ for oblate and prolate spheroids, implying that sedimentation 
in a diatomic gas subject to a background magnetic field can separate 
particles according to shape.} \end{abstract}

\maketitle

\tableofcontents

\section{Introduction}
What makes solving fluid mechanical problems appealing is that there is no bound on the scales at which its outcome can be applied. This universal feature inspired Avron in the late 1990s to study the fluid mechanical applications of odd viscosity in two dimensions \cite{avron1998odd}, after odd viscosity was proposed as a topological quantum phenomenon \cite{avron1995viscosity,levay1995berry} in the wake of the discovery of the quantum Hall effect \cite{Hoyos_2014}.
In the next 20 years, the study of odd viscous flow was ignored, up to some exceptions \cite{PhysRevE.89.043019,PhysRevE.90.063005}, after which odd viscosity was experimentally observed in two completely distinct two-dimensional systems, being graphene under a magnetic field \cite{berdyugin2019measuring} and rotating colloids \cite{soni2019odd}. This discovery sparked countless papers studying odd viscous flow in two dimensions \cite{ganeshan2017odd,abanov2018odd,hosaka2021hydrodynamic,hosaka2023hydrodynamics,hosaka2021nonreciprocal}. It was found that for incompressible flow in two dimensions, odd viscosity is reluctant to reveal itself, prompting subsequent theoretical studies of incompressible three-dimensional flows with an \textit{axis of chirality} that breaks rotational symmetry \cite{olvera,everts2023dissipative,hosaka2021nonreciprocal,Lier_2024,khain2023trading,Khain_2022,markovich2021odd,PhysRevFluids.7.114201}. Remarkably, despite that the study of three-dimensional odd viscous flow only came in full swing in the past five years, first observation of odd viscous flow in three dimensions took place in the 1960s already, when it was observed in the context of the \textit{Senftleben-Beenakker effect}.
\newline 
The Senftleben-Beenakker effect is the phenomenon where gases composed of nonspherical particles, such as diatomic particles, see their transport affected by a background magnetic field, which plays the role of the axis of chirality. In particular, rotation of diatomic particles makes these particles effectively disklike and magnetic moment points along the normal of the disk, so that magnetic field induces precession which affects the nature of the particle collisions \cite{Gorter1938}. As a consequence, the transport coefficients of a diatomic gas are affected by magnetic field \cite{alma990001320080205131}. Whereas Senftleben observed that magnetic field acts to lower parity-even transport \cite{Senftleben1930PhysZ,Beenakker1962,KORVING1967177,KORVING1967198,HULSMAN1972501}, it was later found by Beenakker et al. that a coupling of magnetic field to the collisional nature of rotating diatomic particles can also give rise to odd viscosity \cite{korvingshort,HULSMAN197053},  as well as odd thermal conductivity \cite{PismaZhETF.4.456}, for paramagnetic but also nonparamagnetic molecules, as the latter have a weaker but nonvanishing magnetic moment. Already before this finding, it was predicted on the  basis of symmetry that when a background magnetic field couples to fluid properties in three dimensions, this allows for two distinct odd viscosities to be nonvanishing \cite{HOOYMAN1954355,degroot1984nonequilibrium}. Three things are remarkable about the discovery of odd viscosity in diatomic gases in a magnetic field.
\begin{enumerate}
    \item Odd viscosity was found for a gas of charge neutral particles\footnote{Max von Laue considered it unlikely that odd transport could be observed for fluids composed of charge neutral particles \cite{https://doi.org/10.1002/andp.19354150102}.}.
    \item The nonvanishing of odd viscosity is a signature of the nonsphericity of diatomic particles, whereas parity-even viscous transport is merely reduced by the Senftleben-Beenakker effect.
    \item When magnetic field is small compared to pressure, the two anisotropic odd viscosities are found to have opposite sign. 
\end{enumerate}
It is the third feature of odd viscosity for a diatomic gas in a magnetic field that will be especially important in this work. To better understand this feature, let us consider the fluid equations for conservation of momentum $\rho \mathbf{u}$ and mass density $\rho$
\begin{align}  \label{eq:eomeom}
       \nabla \cdot \bm  \sigma   &  =   \rho  ( \dot{\mathbf{u}}  +   \mathbf{u} \cdot \nabla   \mathbf{u}   ) ~~ , ~~  \dot{\rho}   + \nabla \cdot  ( \rho \mathbf{u}  )  =0    ,  \end{align}
where $\bm  \sigma $ is the stress tensor. We consider a three-dimensional \textit{odd viscous fluid} with an azimuthal symmetry. This fluid conserves angular momentum, which means that $\bm \sigma$ is a symmetric two-tensor. We allow for parity breaking only in conjunction with time-reversal symmetry breaking, as is consistent with the breaking induced by a background magnetic field. Let us define the strain rate as 
\begin{align}
\mathbf{e}= \frac{1}{2} \left(  \nabla \mathbf{u}  + \left( \nabla \mathbf{u} \right)^T  \right)    ~~  , 
\end{align}
and decompose the stress as
\begin{align}
 \bm    \sigma = - \mathbf{1} p  + \bm  \tau ~~ , 
\end{align}
where $p $ is pressure and $ \bm  \tau$ is the viscous stress. Then, the most general viscous stress is given by \cite{HOOYMAN1954355,degroot1984nonequilibrium}
\textcolor{black}{\begin{equation}  \label{eq:cylindrical}
\resizebox{\columnwidth}{!}{$
\begin{pmatrix}
\tau_V\\
\tau_S^{1}\\
\tau_S^{2}\\
\tau_S^{3}\\
\tau_S^{4}\\
\tau_S^{5}
\end{pmatrix}
=
2
\begin{pmatrix}
\zeta_V   & 0 & 0 & \zeta_3   & 0 & 0 \\
0 & \eta_{\perp}  &  2 \gamma_{\perp } & 0 & 0 & 0 \\
0 & - 2 \gamma_{\perp }  & \eta_{\perp}  & 0 & 0 & 0 \\
2 \zeta_3  & 0 & 0 & \eta_3  & 0 & 0 \\
0 & 0 & 0 & 0 & \eta_\parallel   &    - \gamma_{\parallel }  \\
0 & 0 & 0 & 0 &  \gamma_{\parallel }  & \eta_\parallel 
\end{pmatrix}
\begin{pmatrix}
e_V \\
e_S^{1}\\
e_S^{2}\\
e_S^{3}\\
e_S^{4}\\
e_S^{5}
\end{pmatrix}
$} , 
\end{equation} } 
where the labels correspond to contractions of the tensors $\bm \tau$ and $\mathbf{e}$ with the matrices \cite{Khain_2022}
\begin{equation}
\begin{aligned}
\mathbf{V} &=
\begin{pmatrix}
 1 & 0 & 0 \\
 0 & 1 & 0 \\
 0 & 0 & 1
\end{pmatrix} ~~ , ~~ \mathbf{S}^{1} =
\begin{pmatrix}
 1 & 0 & 0 \\
 0 & -1 & 0 \\
 0 & 0 & 0
\end{pmatrix},
\\
\mathbf{S}^{2} &=
\begin{pmatrix}
 0 & 1 & 0 \\
 1 & 0 & 0 \\
 0 & 0 & 0
\end{pmatrix} ~~ , ~~ 
\mathbf{S}^{3}  =
\begin{pmatrix}
 -1  & 0 & 0 \\
 0 & -1  & 0 \\
 0 & 0 & 2 
\end{pmatrix},
\\
\mathbf{S}^{4} &=
\begin{pmatrix}
 0 & 0 & 0 \\
 0 & 0 & 1 \\
 0 & 1 & 0
\end{pmatrix} ~~ , ~~ 
\mathbf{S}^{5} =
\begin{pmatrix}
 0 & 0 & 1 \\
 0 & 0 & 0 \\
 1 & 0 & 0
\end{pmatrix}.
\end{aligned}
\end{equation}
A kinetic theory computation that was experimentally corroborated predicted that the odd viscosities in \eqref{eq:cylindrical} have the form \cite{KNAAP1967643,KaganMaksimov1962}
\textcolor{black}{\begin{subequations}  
\label{eq:allgamma}
    \begin{align}  \label{eq:oddviscoformulae}
    \gamma_{\perp }   &  = -K_1  \left\{ 4 \frac{ \Theta}{1 + 4  \Theta^2 }  +   3 \frac{\Theta}{1 + \Theta^2 }     \right\} ~~  ,    \\ 
       \gamma_{\parallel}   &  = - K_1 \left\{  12  \frac{ \Theta}{1 + 4 \Theta^2 }   -   5 \frac{\Theta}{1 + \Theta^2 }    \right\}  ~~ , 
\end{align}
\end{subequations}} 
with 
\begin{align}
    \Theta = K_2  \frac{H}{p} ~~ , 
\end{align}
where $H$ is the magnetic field and $p$ is the pressure, $K_{1,2}$ are constants that depend on the molecule's magnetic moment as well as the coefficient that gives the coupling between angular momentum and relative velocity in the collision integral. In experiment, it is possible to tune both $p$ and $H$ and the experimentalist is thus enabled to change the sign of $\gamma_\parallel $ by adequately choosing these parameters. 
\newline 
It is a unique feature of odd viscous transport coefficients that they can vary so drastically without violating the Second Law of Thermodynamics, which states that entropy production cannot be negative. For parity-even incompressible Stokes flow, where the viscous four-tensor is symmetric under the exchange of the first and last index pair, nonnegativity of entropy production strongly constrains flow. For example, it guarantees that
\begin{itemize}
    \item The Stokes solution is always the least dissipative incompressible fluid profile satisfying the boundary conditions \cite{Helmholtz1868Discontinuous,Rayleigh01101913,lamb1932hydrodynamics}
    \item Drag on a body $A$ is always less than drag on a body $B$ when $A$ is enclosed by $B$ and the bodies translate at the same speed. This is called \textit{inclusion monotonicity} \cite{kim2005microhydrodynamics,hillpower}.
\end{itemize}
Because the odd viscous tensor does not enter the dissipation rate, its flow cannot be constrained using the nonnegativity of dissipation. Thus there exists no odd viscous equivalent of these powerful theorems.
\newline 
Because the sign flip of odd viscosity happens for only one of the two anisotropic odd viscosities $\gamma_\parallel $, one must separately treat these two odd viscosities when solving for the flow in order to fully appreciate its implications. For this, one cannot rely in previous literature, as all of the many recent works on three-dimensional odd Stokes flow considered a single \quotes{isotropic} odd viscosity \cite{khain2023trading,everts2023dissipative,olvera,Khain_2022,khain2023trading,Lier_2024,hosaka2023lorentz,markovich2021odd,PhysRevResearch.6.L032044,everts2023dissipative,dewit2025nonhermitianwaveturbulence}, and it is therefore what is done in this work. In particular, we study a mechanism by which prolate and oblate spheroids can be separated through sedimentation at low Reynolds number. Throughout, we will assume incompressibility, which means that \eqref{eq:eomeom} thus reduces to
\begin{align}  \label{eq:eomyeah}
       \nabla \cdot \bm \sigma   &   =0  ~~ , ~~ \nabla \cdot \mathbf{u}  =0   .   
\end{align}
Furthermore, incompressibility means that $\zeta_V$ in \eqref{eq:cylindrical} can be discarded. 
\begin{figure}
    \centering
    \tikz[thick]{
        \node (img) at (0,0) {\includegraphics[width=1\linewidth]{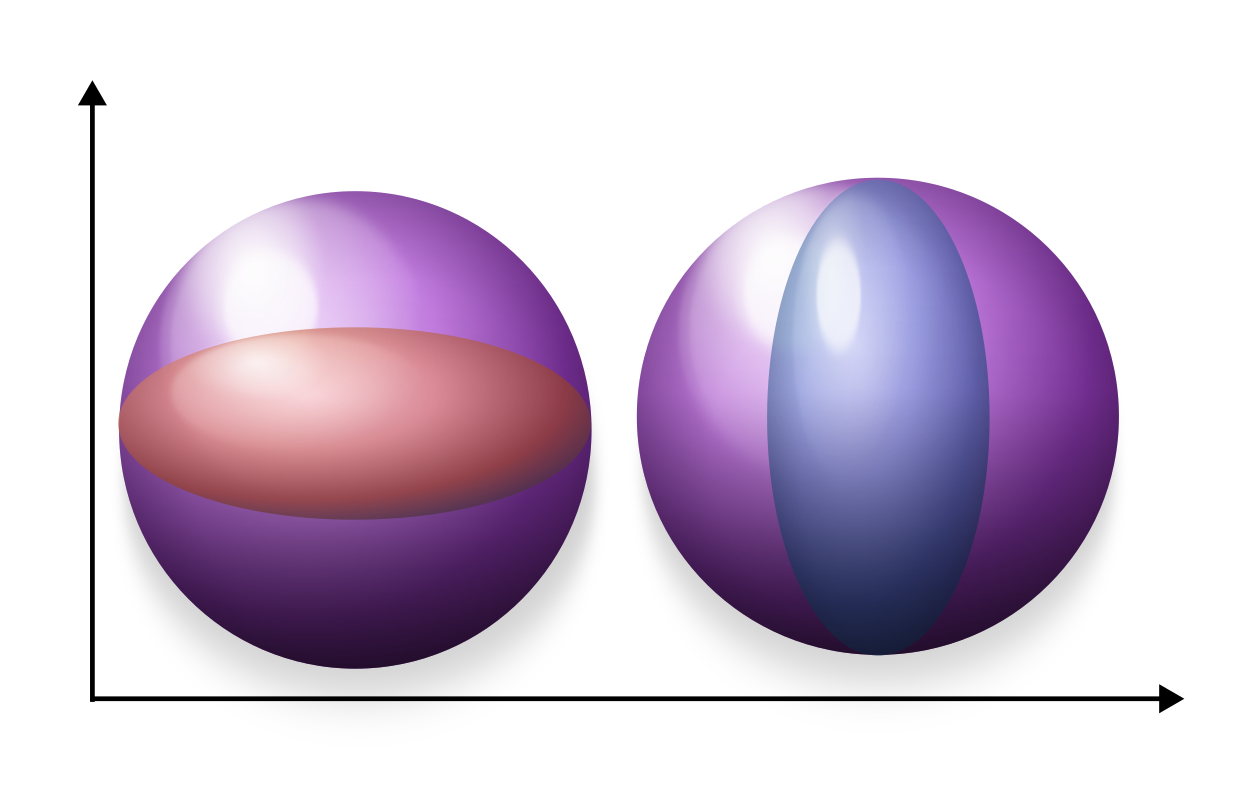}};
   \node[fill=none,scale=1.3] at (0.25 , -5.6+3.2) {$x,y$};
        \node[fill=none,scale=1.3] at (-4, -2+2) {$z$};
         \node at (-1.05, 0.95) {$S_{\sphere }$};
         \node at (2.57, 0.95) {$S_{\sphere }$};
         \node at (2.2, 0.3) {$S_{\prol }$};
         \node at (-1,0.1) {$S_{\obl }$}; }
    \caption{Side view of the oblate (red) and prolate (blue) spheroid enclosed by the reference sphere (purple). The $z$-direction is the symmetry axis along which the magnetic field is pointed.}
    \label{figsimpleimpact1224}
\end{figure}
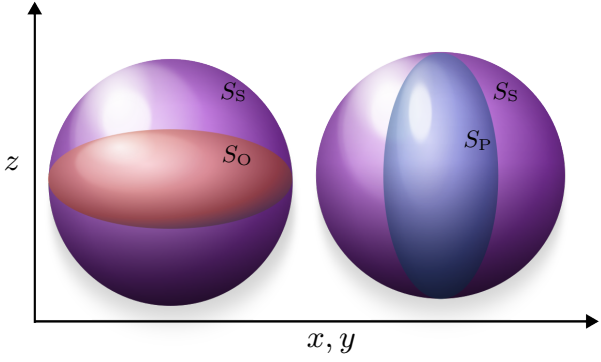
\section{Anisotropic odd lift force on a sphere}
We will assume magnetic contribution to the fluid to be small, and therefore only account for linear parity-odd corrections coming from the magnetic field, which simplifies the parity-even transport in \eqref{eq:cylindrical} to  
\textcolor{black}{\begin{align}
    \eta_{\parallel } = \eta_{\perp  }   = \eta_\sh  ~~ , ~~ \zeta_3  = \eta_3 =0  ~~  . 
\end{align}} 
It is consistent to ignore parity-even magnetic corrections but keep odd viscosity, as we will be computing lift force on a symmetric object which, in order to be nonzero, must be at least linear in the parity-odd magnetic contribution. In contrast, parity-even magnetic corrections can merely produce a subleading correction. Let us introduce the unit vector $ \bm \ell$ which points along the axis of chirality, i.e. the magnetic field. The viscous stress can then be written as 
\begin{align}
\begin{split}
 \bm    \tau  & =   \eta_\sh  \left[ \nabla \mathbf{u}   +  ( \nabla \mathbf{u} )^\text{T}  \right] + 
    \\  &  +   \gamma_{\perp } \left[ \nabla^* \mathbf{u}^{\perp}   +  (  \nabla^* \mathbf{u}^{\perp})^\text{T} +  \nabla^{\perp} \mathbf{u}^{ *}   +  (  \nabla^{\perp} \mathbf{u}^*  )^\text{T}  \right]  + 
    \\  &  +    \gamma_{\parallel } \left[ \nabla^* \mathbf{u}^{\parallel }   +  (  \nabla^* \mathbf{u}^{\parallel })^\text{T} +  \nabla^{\parallel } \mathbf{u}^{ *}   +  (  \nabla^{\parallel }  \mathbf{u}^*)^\text{T}  \right]  ~~ , 
\end{split}
\end{align}
where 
\begin{align}
   \mathbf{u}^{\parallel} =   \bm \ell   \bm \ell \cdot \mathbf{u}  ~~ , ~~  \mathbf{u}^{\perp} =    ( \mathbf{1} - \bm  \ell   \bm \ell ) \cdot \mathbf{u} ~~ . 
\end{align}
and $\mathbf{u}^* =  \bm \varepsilon \cdot \mathbf{u}$, where $\bm \varepsilon$ is the two-dimensional anisotropic Levi-Civita symbol, which is related to the three dimensional Levi-Civita symbol $\bm \varepsilon^{(3)}$ as $\bm \varepsilon = \bm \ell \cdot  \bm \varepsilon^{(3)}$. Using \eqref{eq:eomyeah}, we find the Stokes equation
\begin{align}  \label{eq:Stokesequation}
\begin{split}
       &  \eta_\sh  \Delta \mathbf{u}    +   \left[  \gamma_{\perp } \Delta_\perp +  \gamma_{\parallel} \Delta_\parallel      \right] \mathbf{u}^* \\  &  +  ( \gamma_{\parallel} - \gamma_{\perp } )  \left[  \nabla^* \nabla \cdot \mathbf{u}^\parallel  +  \nabla^\parallel  \nabla \cdot \mathbf{u}^*  \right]  = \nabla    \tilde p  \,  ,  
        \end{split}
\end{align}
where $\Delta$ is the Laplacian operator and we introduced the \quotes{modified pressure}\cite{ganeshan2017odd}
\begin{align}
   \tilde p  =    p -  \gamma_{\perp }  \nabla \cdot \mathbf{u}^*  ~ .  
\end{align}
We consider odd viscosity to be a small magnetic correction to the isotropic parity-even flow and thus proceed perturbatively. For this, let us take $\gamma_{\perp , \parallel } \rightarrow  \epsilon \gamma_{\perp , \parallel }  $, where $\epsilon$ is a dimensionless bookkeeping parameter. Correspondingly, we expand the fluid velocity as 
\begin{align}
\mathbf{u}=\mathbf{u}^{(\even)}+\epsilon \mathbf{u}^{(\odd)}+\mathcal{O} (\epsilon^2 ) ~~ .  
\end{align}
\textcolor{black}{We wish to obtain the odd viscous contributions to the force for an object that moves at fixed velocity $\mathbf{U}$, which means we must similarly expand the force as 
\begin{align}
\mathbf{F}=\mathbf{F}^{(\even)}+\epsilon \mathbf{F}^{(\odd)}+\mathcal{O} (\epsilon^2 ) ~~  . 
\end{align}  } 
To solve for the flow, we go to reciprocal space. When the fluid experiences a singular force $\mathbf{F}$ at $\mathbf{x} =0 $, the leading-order reciprocal-space Stokeslet solution is given by
\begin{equation}  \label{eq:stokeslet}
\mathbf{ \tilde u}^{(\even)}= \mathbf{ \tilde G}^{(\even)} \cdot \mathbf{F}^{(\even)} 
 ~~ , ~~   \mathbf{ \tilde G}^{(\even)} =\frac{1}{\eta_\sh }\,\frac{  \mathbf{P} }{k^2},
\end{equation}
where we introduced the transverse projector $\mathbf{P} = \mathbf{1} -\mathbf{ \hat  k } \mathbf{ \hat  k }$. \eqref{eq:Stokesequation} at $O(\epsilon)$ in
$\mathbf{k}$-space can be written as
\begin{align}  \label{eq:nop1equation}
\begin{split}
 - \eta_\sh  k^2\, \mathbf{ \tilde u}^{(\odd)}
  & = \,
\left[ \gamma_\perp k_\perp^2   +\gamma_\parallel k_\parallel^2   \right]  \mathbf{P} \cdot \mathbf{ \tilde u}^{ * (\even)}    \\ 
&     + (\gamma_\parallel - \gamma_\perp ) \mathbf{P} \cdot   \left[  \mathbf{k}^*  \mathbf{k}^\parallel  \cdot  \mathbf{ \tilde u}^{(\even)}  +  \mathbf{k}^\parallel  \mathbf{k} \cdot  \mathbf{ \tilde u}^{*  (\even) }   \right] ~~  . 
\end{split}
\end{align}
Again considering the singular force $\mathbf{F}$, we find the subleading solution 
\begin{align}  \label{eq:combinedthing}
  \mathbf{ \tilde u}^{(\odd)}  & =  \mathbf{\tilde G}^{(\odd)} \cdot \, \mathbf{F}^{(\even)}  + \mathbf{\tilde G}^{(\even)} \cdot \, \mathbf{F}^{(\odd)}  , 
\end{align}
To see the contents of $\mathbf{ \tilde G}^{(\odd)}$, let us decompose it into an isotropic and anisotropic part as
\begin{align}
    \mathbf{\tilde G}^{(\odd)}
= \mathbf{\tilde G}^{(\odd),\text{ISO}}
 + \mathbf{\tilde G}^{(\odd),\text{ANI}} ~~ , 
\end{align}
where 
\begin{subequations}
    \begin{align}  \label{eq:isoiso}
  \mathbf{\tilde G}^{(\odd),\text{ISO}}   & =   - 
\frac{\gamma_\perp }{ \eta_s^2 k^2}\,
\mathbf{P}  \cdot  \bm \varepsilon  \cdot  \mathbf{P}  ~~ ,  \\ 
\begin{split}
   \mathbf{\tilde G}^{(\odd),\text{ANI}}   &  =   - 
\frac{(\gamma_\parallel  - \gamma_{\perp } )  }{\eta_s^2 k^4}\,  \Bigg[  k_\parallel^2 
 \mathbf{P}  \cdot  \bm \varepsilon  \cdot  \mathbf{P}    \\ 
 & +    \mathbf k^*  \mathbf k^\parallel  \cdot  \mathbf{P}    -   \mathbf{P}  \cdot \mathbf k^\parallel  \mathbf k^*       \Bigg]   ~~ . 
\end{split}
\end{align}
\end{subequations}
When $\gamma_\parallel= \gamma_{\perp }$, only $ \mathbf{\tilde G}^{(\odd),\text{ISO}} $ contributes. This isotropic case has been worked out in detail in previous works \cite{olvera,everts2023dissipative,markovich2021odd,hosaka2023lorentz,khain2023trading,Lier_2024,deWit2024,dewit2025nonhermitianwaveturbulence,PhysRevLett.133.144002}. In App.~\ref{app:IFTsection}, we follow the singularity method \cite{happel1983low,kim2005microhydrodynamics,everts2023dissipative,meissneroszer2025exactresultsdissipationsteady} to solve the problem of flow around the no-slip sphere. Specifically, this method involves an inverse Fourier transform of $ \mathbf{\tilde G}^{(\odd),\text{ISO}} $ combined with the spherical Bessel function that is consistent with the multipole expansion of spherical flow. For this method to work it is required that force density is uniformly distributed around the obstacle \cite{levine2001response,weisenborn1984oseen,meissneroszer2025exactresultsdissipationsteady}. After having obtained the corresponding fluid profile, we can verify that the solution is self-consistent and the assumption of uniform distribution is valid. Having confirmed the validity of the obtained velocity profile, the pressure can be obtained simply by performing an indefinite integral of \eqref{eq:Stokesequation}. With the velocity and pressure profile obtained, the force is found to be given by
\begin{align}  \label{eq:Forcequation}
    \mathbf{F}_\sphere  =  - 6 \pi a \eta_s  \left(\mathbf{1}   +  \frac{1}{10} \left(4 \gamma_\perp  +   \gamma_\parallel  \right)    \bm \varepsilon   \right) \cdot  \mathbf{U}   ~~ , 
\end{align}
where 
\begin{align}
    \mathbf{F}_\sphere = \int_{S_\sphere} d \mathbf{n} \cdot  \bm \sigma ~~ .  
\end{align}
\begin{figure}
    \centering
    \tikz[thick]{
        \node (img) at (0,0) {\includegraphics[width=1\linewidth]{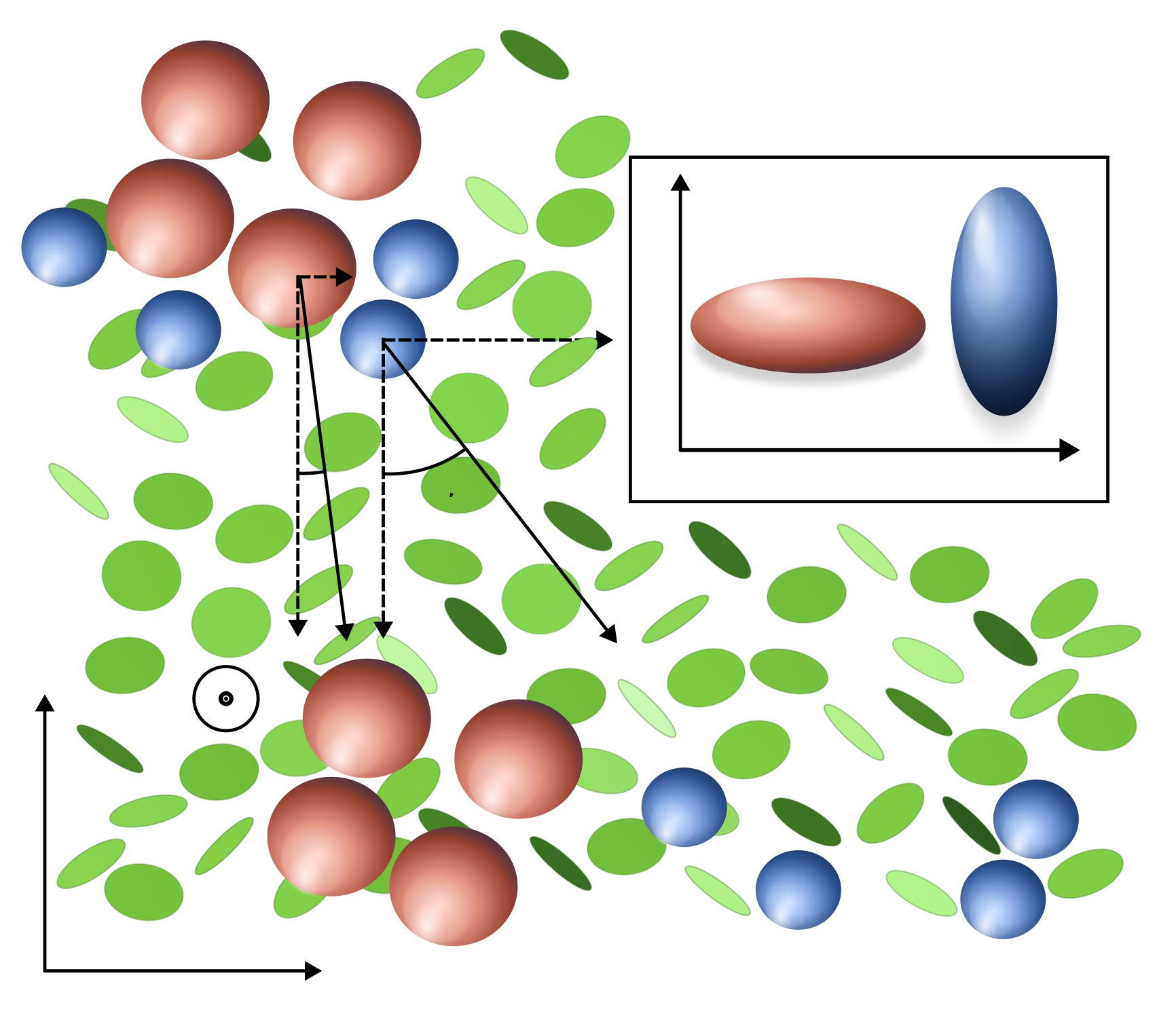}};
        \node at (-3, -3.6) {$x$};
        \node at (-0.8, 2.6) {$t=0$};
                \node at (0.2, -2.9) {$t \rightarrow \infty $};
        \node at (-4.2, -2.5) {$y$};
   \node at (-3+4.7, -3.6+3.85) {$x,y$};
        \node at (-4.2+4.7, -2.5+3.85) {$z$};
         \node at (-5.2+3.65+0.3, -2.5+3) {$\psi_{\prol }$};
         \node at (-4.2+1.5+0.3, -2.5+2.8) {$\psi_{\obl }$};
        \node at (-4.2+4.7, -2.5+3.85) {$z$};
        \node at (-6.2+3.05, -2.5+0.9) {$\mathbf{H}$};    }
    \caption{Top view of oblate (red) and prolate (blue) spheroids sedimenting in a gas of rotating diatomic particles (green) in such a way that a cluster of oblate and prolate spheroids is sorted. The magnetic field $\mathbf{H}$ is pointed out of plane. Inset: side view of the oblate and prolate spheroid.}
    \label{figsimpleimpact}
\end{figure}
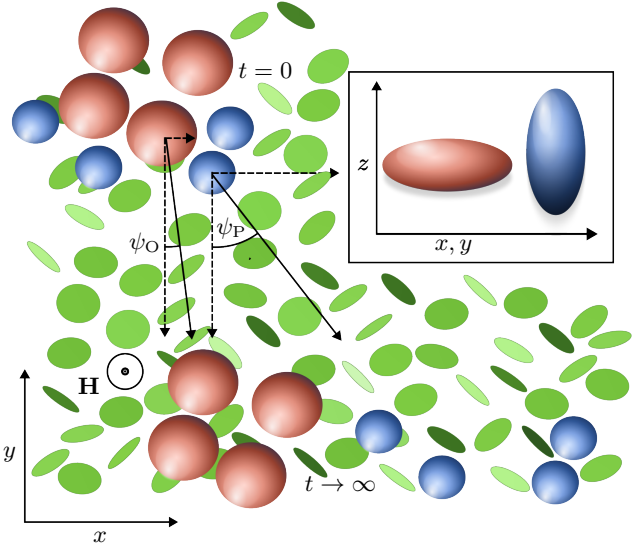
\section{Perturbative anisotropic odd lift force on a spheroid}
Having found the solution for the sphere, we consider the problem of the spheroid. It is possible to solve for the problem of Stokes flow around a spheroid exactly as has been done for the case where there is only shear viscosity \cite{lamb1932hydrodynamics,10.1098/rspa.1922.0078,Sampson1891,Payne_Pell_1960,Oberbeck+1876+62+80,jeffery1922motion}, but to generalize this solution for the case where the viscous differential operator has a parity odd contribution and is anisotropic is challenging, especially considering that anisotropy forbids the use of a stream function. Therefore, we instead follow the perturbative approach valid for a slightly deformed sphere \cite{BRENNER1964519,happel1983low} and only nonperturbatively solve for spheroidal flow for the isotropically odd viscous case. For an oblate spheroid, the radius on the spheroid boundary is given by
\begin{align}
r =  \sqrt{a^2 \sin^2 (\theta)  + b^2 \cos^2 (\theta) }  ~~  \text{on $S_{\obl}$}  ~~  , 
\end{align}
where $\theta$ is the polar angle and $b  \leq  a $. Assuming a slightly deformed sphere, we expand as
\begin{align}
r = a\left( 1 -   \kappa   \cos^2(\theta)  \right)  + \mathcal{O} (\kappa^2 )  ~~  \text{on $S_{\obl}$}  ~~  , 
\end{align}
where $\kappa$ is given by 
\begin{align}  \label{eq:kappaexpressionoblate}
 \kappa =  1 -  \frac{b}{a}     ~~ . 
\end{align}
To see the effect of ellipticity, we Taylor expand with respect to $r=a$ as 
\begin{align}
     \mathbf{u} \big|_{S_{\obl}}  =  \mathbf{u} \big|_{r = a }    +    \sum_{j=1 } \frac{  \left( -  \cos^2(\theta)  a  \kappa \right)^j }{j ! }  \frac{\partial^{(j) }  \mathbf{u}}{\partial  r^{(j)}  }  \big|_{r = a } ~~ . 
\end{align}
We expand as
\begin{align}
    \mathbf{u}  =   \mathbf{u}^{(0) } +   \kappa  \mathbf{u}^{(1 ) }  +  \mathcal{O}  ( \kappa^2  ) ~~  , 
\end{align}
where $ \mathbf{u}^{(0) }$ is the spherical solution. Since $\mathbf{u} \big|_{S_{\obl}} =0 $, the first order spheroidal correction is then related to the spherical solution as
\begin{align} \label{eq:correct}
     \mathbf{u}^{(1)} \big|_{r = a }   =    a    \cos^2(\theta)    \frac{\partial    \mathbf{u}^{(0)}}{\partial   r   }  \big|_{r = a } ~~ . 
\end{align}
To find the force correction due to ellipticity, one can view \eqref{eq:correct} as an effective slip velocity on a reference sphere which encloses the original spheroid (see Fig.~\ref{figsimpleimpact1224}). Then, one considers the Lorentz reciprocal relation \cite{lorentzoriginal,masoud2019reciprocal}
\begin{align}  \label{eq:LRTrelation}
 \bm \sigma' :   \nabla \mathbf{u}  =  \bm \sigma :   \nabla \mathbf{u}'   ~~ , 
\end{align}
where the $\bm \sigma^{\prime}$ and $\mathbf{u}'$ correspond to an auxiliary fluid system where the sphere is moving with velocity $\mathbf{U}'$, the reference sphere is slip-free, and the odd viscosities have their sign flipped compared to the unprimed fluid system \cite{hosaka2023lorentz}, keeping the rest of the stress tensor identical. That is, we have
\begin{align}
    \bm \sigma' = - \mathbf{1} p'    + \bm \tau' ~~ , 
\end{align}
with
\begin{align}
\begin{split}
  \bm   \tau'  & =   \eta_\sh  \left[ \nabla \mathbf{u}^{ \prime}   +  ( \nabla \mathbf{u}^{ \prime} )^\text{T}  \right] + 
    \\  &  -    \gamma_{\perp } \left[ \nabla^* \mathbf{u}^{ \prime\perp}   +  (  \nabla^* \mathbf{u}^{ \prime\perp})^\text{T} +  \nabla^{\perp} \mathbf{u}^{ \prime *}   +  ( \nabla^{\perp} \mathbf{u}^{ \prime *}  )^\text{T}  \right]  + 
    \\  &  -    \gamma_{\parallel  } \left[ \nabla^* \mathbf{u}^{ \prime\parallel }   +  (  \nabla^* \mathbf{u}^{ \prime\parallel })^\text{T} +  \nabla^{\parallel } \mathbf{u}^{ \prime *}   +  (  \nabla^{\parallel } \mathbf{u}^{ \prime *}  )^\text{T}  \right]  ~~ . 
\end{split}
\end{align}
Integrating \eqref{eq:LRTrelation} over the volume surrounding the reference sphere $S_\sphere$, we find
\begin{align}  \label{eq:LRTequation}
    \mathbf{F}_{\obl}  \cdot \mathbf{U}' =   \mathbf{F}' \cdot \mathbf{U}  + \int_{S_\sphere} d \mathbf{n} \cdot  \bm \sigma^{\prime }  \cdot   \mathbf{u}^{(1)}   + \mathcal{O} ( \kappa^2 )   ~~ ,  
    \end{align}
    where we introduced the forces on the reference sphere given by
    \begin{align}
         \mathbf{F}_{\obl}   = \int_{S_\sphere} d \mathbf{n} \cdot  \bm \sigma     ~~ , ~~ \mathbf{F}' = \int_{S_\sphere} d \mathbf{n} \cdot  \bm \sigma'  
    \end{align}
    To extract the leading order parity-even force on an oblate spheroid moving orthogonally to $\bm \ell $, we take $\mathbf{U} \cdot \bm  \ell  =0 $ and $\mathbf{U}^{\prime} = \mathbf{U}$ and plug \eqref{eq:correct} into \eqref{eq:LRTequation}, which yields
\begin{subequations}
\label{eq:oblatething}
        \begin{align}
        \mathbf{F}_{\obl} \cdot \mathbf{U} =     \mathbf{F}' \cdot \mathbf{U}  +  \frac{12 \kappa }{5} \pi a U^2  + \mathcal{O} ( \epsilon^2 ,\kappa^2 ) ~~ .   
    \end{align}
To extract the odd force, we take $\mathbf{U}^{\prime} = -\mathbf{U}^*$, which yields
 \begin{align}  \label{eq:oddforce}
 \begin{split}
     &        -  \mathbf{F}_{\obl} \cdot \mathbf{U}^*  \\  & =    \mathbf{F}' \cdot \mathbf{U} - \frac{3 \pi \kappa \epsilon }{175} a U^2   \left(  52    \gamma_\perp  -17   \gamma_\parallel      \right) + \mathcal{O} ( \epsilon^3 ,\kappa^2 )  .    \end{split}
    \end{align}
    \end{subequations}
We can repeat the same process for prolate spheroids, for which we have
\begin{align}
r =  \sqrt{b^2 \sin^2 (\theta)  + a^2 \cos^2 (\theta) }  ~~  \text{on $S_{\prol}$}  ~~  , 
\end{align}
which means that 
\begin{align} \label{eq:correct12}
     \mathbf{u}^{(1)} \big|_{r = a }   =    a   \kappa  \sin^2(\theta)    \frac{\partial    \mathbf{u}^{(0)}}{\partial   r   }  \big|_{r = a } ~~ . 
\end{align}
    We find for the parity-even force
    \begin{subequations} 
     \label{eq:prolatething}
            \begin{align}
        \mathbf{F}_{\prol} \cdot \mathbf{U} =     \mathbf{F}' \cdot \mathbf{U}  +  \frac{18 \kappa }{5} \pi a U^2  + \mathcal{O} ( \epsilon^2 ,\kappa^2 ) ~~ .   
    \end{align}
For odd lift force we find
 \begin{align}
 \begin{split}
         &     -    \mathbf{F}_{\prol} \cdot \mathbf{U}^* \\  & =   \mathbf{F}' \cdot  \mathbf{U}  - \frac{12 \pi \kappa \epsilon}{175} a U^2   \left(  22    \gamma_\perp   + 13   \gamma_\parallel      \right) + \mathcal{O} ( \epsilon^3 ,\kappa^2 )  .   
             \end{split}
    \end{align}
        \end{subequations}
        We learn from \eqref{eq:oblatething} and \eqref{eq:prolatething} that the drag force, which at leading order is only affected by shear viscosity, deformations of the reference sphere that act to decrease its size can only lower this force, which is consistent with inclusion monotonicity. For lift force, we find that for three instances, the shrinking deformation with respect to the reference sphere lowers the force. In \eqref{eq:oddforce}, we see a crucial exception, where the leading order $\kappa$ correction proportional to $\gamma_{\parallel}$ \textit{increases} lift, despite the oblate spheroid being smaller than the reference sphere. This phenomenon is possible for odd viscosity as odd viscosity is nondissipative and thus its corresponding flow is not subject to inclusion monotonicity. 
        \section{Sedimentation of spheroids in anisotropic odd viscosity}
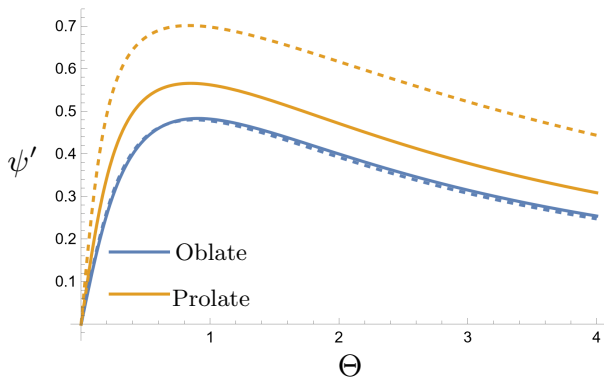
\begin{figure}
    \centering
    \tikz[thick]{
        \node (img) at (0,0) {\includegraphics[width=0.85\linewidth]{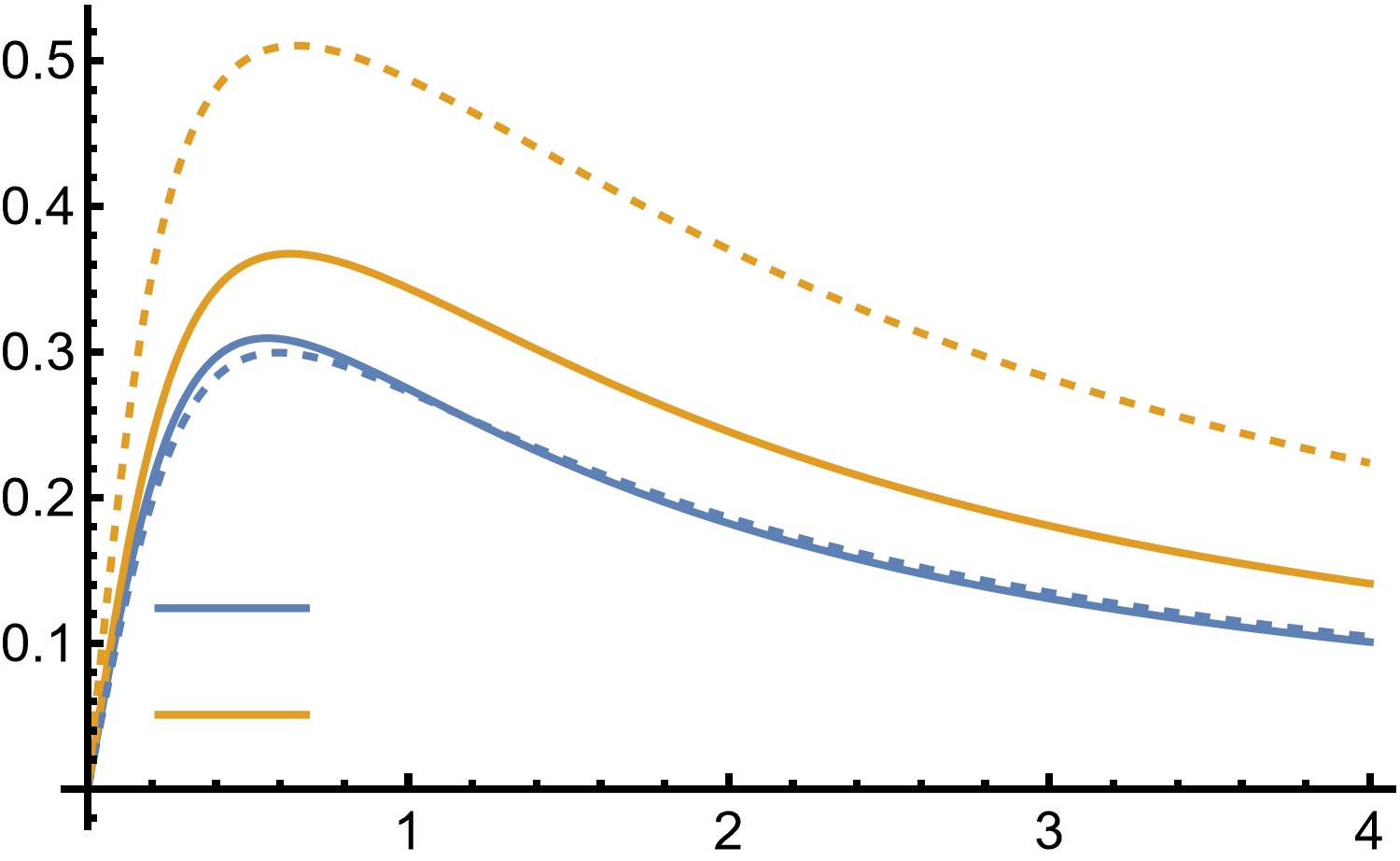}};
   \node[fill=none,scale=1.3] at (0.25 , -5.2+2.7) {$\Theta$};
        \node[fill=none,scale=1.3] at (-4.1, -2+2.2) {$  \psi'$} ;
         \node at (-1.5, -0.9) {$\text{Oblate}$};
         \node at (-1.5, -1.5) {$\text{Prolate}$}; }
    \caption{Rescaled Hall angle $\psi' =  -  \psi / (\pi /2 )$ as given by \eqref{eq:thetavalue}, for the oblate and prolate spheroid. The solid (dashed) lines correspond to $\kappa =0.5$ ($\kappa =1$). We took $K_2' = 0.5 $. }
    \label{figsimpleimpact122}
\end{figure}

        To exploit this finding, let us now consider the scenario where a mixture of oblate and prolate spheroids that undergoes sedimentation. We assume that the axis of symmetry of all the spheroids is aligned with the magnetic field, which can be enforced when, like the diatomic particles, the spheroids have a magnetic moment which is pointed along this axis. Nevertheless, it will be hard for particles moving in a gas to achieve terminal velocity so that sedimentation can take place and therefore an alternative experimental setup could be considered where the particles are kept fixed. Proceeding, we define the force as
\begin{align}
    \mathbf{F}_{\obl , \prol} = -  \left( \zeta^{ ( \even )}_{\obl , \prol} \mathbf{1}  + \epsilon  \zeta^{(\odd )}_{\obl , \prol}  \bm \varepsilon  \right) \cdot \mathbf{U} + \mathcal{O} ( \kappa^2 , \epsilon^2 )   ~~ . 
\end{align}
For $\mathbf{U} \cdot \bm \ell =0  $, we can extract from \eqref{eq:Forcequation}, \eqref{eq:oblatething} and \eqref{eq:prolatething} the values 
\begin{subequations}
  \begin{align}
    \zeta^{ ( \even )}_{\obl }  & = 6 \pi a \eta_\sh \left[ 1   -    \frac{2  \kappa }{5}     \right]   ~~ , \\
    \begin{split}
   \zeta^{ ( \odd )}_{\obl }  & =  \frac{3}{5} \pi a  \bigg[   \left(4 \gamma_\perp  +   \gamma_\parallel  \right)           - \frac{    \kappa }{35} \left(  52    \gamma_\perp   -  17    \gamma_\parallel      \right)     \bigg]   ~~ ,    \end{split}
      \\ 
    \zeta^{ ( \even )}_{ \prol}  & = 6 \pi a \eta_\sh  \left[ 1   -    \frac{3  \kappa }{5}   \right]  ~~ , \\
    \begin{split}
                  \zeta^{ ( \odd )}_{\prol }  & =  \frac{3}{5} \pi a   \bigg[  \left(4 \gamma_\perp  +   \gamma_\parallel  \right)  
          - \frac{4   }{35}   \kappa  \left(  22    \gamma_\perp   + 13   \gamma_\parallel      \right)    \bigg]   ~~ .
              \end{split}
\end{align}  
\end{subequations}
The Hall angle that the sedimenting oblate and prolate spheroids make is then given by
\begin{align}  \label{eq:Hallangles}
  \psi_{ \obl , \prol }    =  \tan^{-1} \left(\frac{   \zeta^{(\odd)}_{\obl , \prol} }{ \zeta^{(\even )}_{\obl , \prol} }  \right) + \mathcal{O} ( \kappa^2 , \epsilon^3 )  ~~ . 
\end{align}
We can plug in \eqref{eq:allgamma} to find
  \textcolor{black}{ \begin{align}  \label{eq:thetavalue}
    \begin{split}
             \psi_{ \obl  }    &   =     \tan ^{-1}\left(  K_2' \frac{\Theta ^3 (1960-968 \kappa )-245 \Theta  (\kappa -5)}{140 \left(\Theta ^2+1\right) \left(4 \Theta ^2+1\right) (2 \kappa -5)}  \right)   \\  & 
     + \mathcal{O} ( \kappa^2 , \epsilon^3 )  ~~ , \\ 
      \psi_{ \prol }    &   =         \tan^{-1} \left(   K_2'  \frac{8 \Theta ^3 (245-31 \kappa )-245 \Theta  (\kappa -5)}{140 \left(\Theta ^2+1\right) \left(4 \Theta ^2+1\right) (3 \kappa -5)}  \right)  \\  &   + \mathcal{O} ( \kappa^2 , \epsilon^3 ) ~~ ,  
          \end{split}
    \end{align}  } 
    where $K'_2 =  \epsilon K_2 / \eta_\sh $. In Fig \ref{figsimpleimpact122}, we plot \eqref{eq:thetavalue} as a function of $\Theta$ for $\kappa = \{  0.5 ,1  \} $.  Note that $\kappa=1$ corresponds to the extremal case where the oblate spheroid is perfectly flat and the prolate spheroid is perfectly thin. In this limit, the expression in \eqref{eq:thetavalue}, which is perturbative in $\kappa$, is not guaranteed to be a valid description. The limit $\kappa \rightarrow 1 $ is discussed further in Sec.~\ref{sec:finalsection}. \textcolor{black}{When the Hall angles corresponding to the prolate and oblate spheroid differ, the prolate and oblate spheroids can be separated through sedimentation, as shown in Fig.~\ref{figsimpleimpact}, where the difference in Hall angles is exaggerated.} In reality, the difference in Hall angle between oblate and prolate spheroids will be smaller and one needs to consider a larger sedimentation domain to see the shape-based sorting effect.  

\section{Nonperturbative lift force on an oblate spheroid}
\label{sec:finalsection}
\begin{figure}
    \centering
    \tikz[thick]{
        \node (img) at (0,0) {\includegraphics[width=0.85\linewidth]{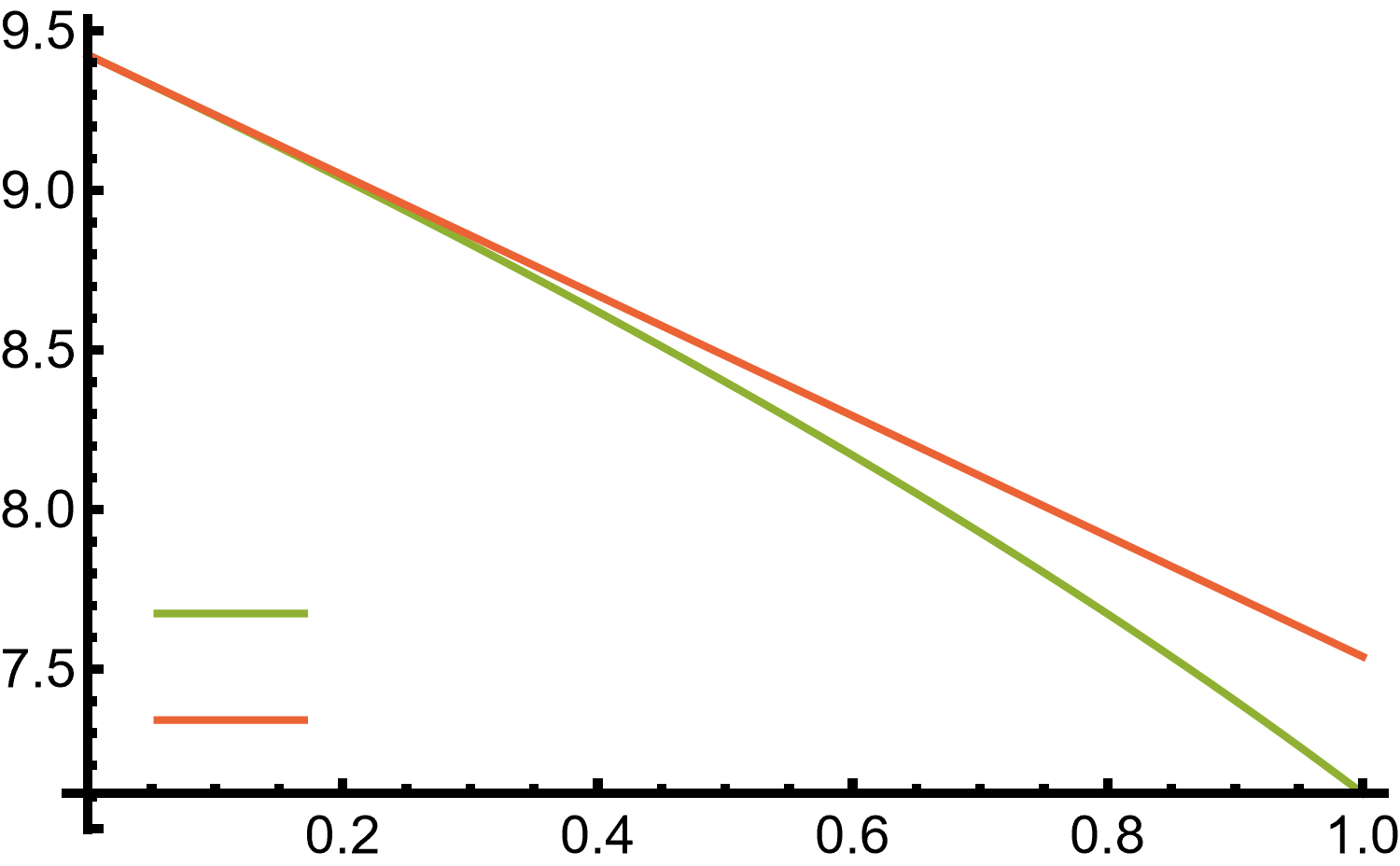}};
   \node[fill=none,scale=1.3] at (0.25 , -5.2+2.7) {\textcolor{black}{$\kappa$}};
        \node[fill=none,scale=1.3] at (-4.5, -2+2.1) {\textcolor{black}{$ \zeta^{(\odd )}_{\obl} / ( \gamma a )   $ } } ;
         \node at (-1.3, -0.9) {$\text{Nonlinear}$};
         \node at (-1.5, -1.5) {$\text{Linear}$}; }
\caption{\textcolor{black}{Comparison between the exact result \eqref{eq:bigresult} and its
linear-in-$\kappa$ approximation \eqref{eq:Xthing} for the normalized
oblate lift coefficient $\zeta^{(\odd)}_{\obl}/(\gamma a)$, plotted as a
function of $\kappa=1-b/a$ over $0\leq\kappa\leq1$. Since both expressions
are proportional to $\gamma$, the normalization removes the dependence on
$\gamma$.}}
    \label{figsimpleimpact1223}
\end{figure}

As a final consistency check, we consider the case where the spheroidal boundary conditions are solved nonperturbatively. As this nonperturbative computation is very tedious and we are reliant on the uniformity of the force density which is not guaranteed, we restrict to the case of a single isotropic odd viscosity $\gamma $. This case has as its advantage that the central computation is identical to that of isotropic shear viscous flow and is thus easy to verify by checking the literature, as is done in App.~\ref{eq:nonperturb}. For the lift force, we find
\begin{align}
    \mathbf{F}^{(\odd  )}_{\obl}   =  -    \zeta^{(\odd )}_{\obl} \bm \varepsilon  \cdot \mathbf{U}      ~~ ,  
\end{align}
with 
\begin{align} \label{eq:bigresult}
  \zeta^{(\odd )}_{\obl}  =     \frac{ 32 \pi  \gamma   (a^2-b^2)^{3/2} \left(a^2 \cot ^{-1}\left(\frac{b}{\sqrt{a^2-b^2}}\right)   - b \sqrt{a^2-b^2} \right)}{\left(\left(2 b^2-3 a^2\right) \cot ^{-1}\left(\frac{b}{\sqrt{a^2-b^2}}\right)+b \sqrt{a^2-b^2}\right)^2}  . 
\end{align}
We can expand \eqref{eq:bigresult} as 
\begin{align}  \label{eq:Xthing}
   \zeta^{(\odd )}_{\obl}   &  =  \pi  a  \gamma    \left(  3 -\frac{3 \kappa  }{5} \right) + \mathcal{O} (\kappa^2 , \epsilon^3  )   ~~  , 
\end{align}
and find that this nonperturbative result is at subleading order consistent with what was found using the Lorentz reciprocal theorem provided one considers also there the isotropic limit for the odd viscosities. Finally, that \eqref{eq:bigresult} is nonperturbative in $\kappa$ allows us to take the limit of an infinitely thin disk \cite{happel1983low,happel1983low} for which we find the lift force coefficient to be 
\begin{align}  \label{eq:extremal}
    \lim_{b \rightarrow 0 } \zeta^{(\odd )}_{\obl}   =    \frac{64 \gamma   }{9}  a      ~~ .  
\end{align}
\textcolor{black}{We can compare the exact result with the truncated expression of \eqref{eq:Xthing}. This is done in Fig.~\ref{figsimpleimpact1223}. It turns out that the truncated expression has an error of only $6\%$ for the extreme case of $\kappa \rightarrow 1 $, which corresponds to $b=0$. This is the same percentage as found for oblate spheroidal drag \cite{happel1983low}. From this we can conclude that also for odd viscous flow, when the spheroidal deformation is accounted for perturbatively for the force computation, the predictive power is great even when the spheroid is far from a slightly deformed sphere.  } 

\section{Discussion}
In this work, we studied three-dimensional odd viscous flow as it arises due to the Senftleben-Beenakker effect. We showed that because the Senftleben-Beenakker effect allows for the two odd viscosities allowed by symmetry to have opposite sign, a diatomic gas in a magnetic field can serve to sort prolate and oblate spheroids. To show this, we solved the problem of incompressible flow around a sphere or spheroid at low Reynolds number, and found that for the longitudinal odd viscosity, shrinking the equatorial radius with respect to that of a sphere can act to \textit{enhance} lift force, an effect that has no shear viscous equivalent, as drag force has to decrease for a shrinking object in accordance with inclusion monotonicity \cite{hillpower}. The consequence of this anisotropic odd viscous phenomenon is that under sedimentation in the direction orthogonal to the magnetic field, a spheroid with its axis of symmetry aligned with the magnetic field will display a very different Hall angle depending on whether it is oblate or prolate. This Hall angle difference arises regardless of the magnitude of the prolate and oblate spheroid, as both lift and drag scale in the same way with the magnitude of the object.
\newline 
Although diatomic gases are commonplace, there are a number of issues with experimentally realizing this sorting effect. Firstly, for the setup to retain its azimuthal symmetry, the spheroids need to have their axis of symmetry aligned with the magnetic field, which can happen when the spheroids have a magnetic moment aligned with the magnetic field. Secondly, the obstacles need to be micron sized, so that one is safely in the Stokes regime. Finally, the gas needs to be sufficiently large and dense, so that the particles can sediment at terminal velocity, although the sorting effect likely already takes place before terminal velocity has been attained. 
\newline 
\textcolor{black}{In future work, it would be interesting to explore the interplay of particle shape with odd viscous flow in different contexts. For example, when a sphere moves along the axis of chirality in odd viscous Stokes flow, torque is zero \cite{Khain_2022,khain2023trading,Lier_2024} due to fore-aft symmetry of the flow \cite{vandyke1975perturbation,herron1975sedimentation,Auregan2023}. An open question is whether an egg-shaped particle that breaks fore-aft symmetry can cause torque to become nonzero for particle motion along the axis of chirality of an odd viscous fluid.}

\section{Acknowledgements}
We thank Joseph E. Avron for useful discussions. \textcolor{black}{This work was funded by the NWA ORC programme Emergence at All Scales.}

\onecolumngrid

\appendix

\section{Singularity method for anisotropic odd viscous flow around a sphere}
\label{app:IFTsection}
We start from the Green's function
\begin{equation} \label{eq:startequation}
   \mathbf{\tilde G}^{(\odd),\text{ANI}}_{ij}     =   - 
\frac{(\gamma_\parallel  - \gamma_{\perp } )  }{\eta_s^2 k^4}\,  \Bigg[  k_\parallel^2 
\mathbf P  \cdot  \bm \varepsilon  \cdot  \mathbf P      +   \mathbf P  \cdot  \mathbf k^*  \mathbf k^\parallel  \cdot  \mathbf P    -   \mathbf P  \cdot \mathbf k^\parallel  \mathbf k^*    \cdot  \mathbf P      \Bigg] . 
\end{equation}
Using the identity \cite{olvera}
\begin{align}
\mathbf P  \cdot  \bm \varepsilon  \cdot  \mathbf P = \bm \varepsilon^{(3)} \cdot \frac{  \mathbf{k}   \mathbf{k}  }{k^2 }  \cdot \bm \ell   ~~ , 
\end{align}
where $ \bm \varepsilon^{(3)}$ is the three-dimensional Levi-Civita symbol, \eqref{eq:startequation} can be written as
\begin{equation}  \label{eq:rewrite123}
\tilde{\mathbf{G}}^{ (\odd),\text{ANI}} (\mathbf{k})
=  -   
\frac{\gamma_\parallel - \gamma_\perp}{\eta_s^2}  \left(  \bm \ell  \bm \ell \cdot  \mathbf{A} \cdot \bm \varepsilon   + \bm \varepsilon \cdot  \mathbf{A} \cdot \bm \ell \bm \ell    +    \bm \varepsilon^{(3)}  \cdot \mathbf{B} \cdot  \mathbf \bm \ell - \mathbf{B} \cdot \bm \varepsilon   - \bm \varepsilon \cdot  \mathbf{B}    \right) 
\end{equation}
with 
\begin{align}
  \mathbf{A}  (\mathbf{k}) =      \frac{\mathbf k      \mathbf k  }{k^4 } ~~ , ~~   \mathbf{B}  (\mathbf{k}) =      \mathbf k      \mathbf k    \,\frac{ k_\parallel^2}{k^6}        ~~ . 
\end{align}
It follows from \eqref{eq:rewrite123} that we only need to compute the inverse Fourier transforms
\begin{align}
\label{eq:thestartingpoint2}
\mathcal{F}^{-1}\!\left[ f_{\sphere}\, \mathbf{A}     \right](\mathbf{r})
= \int \frac{d^3 \mathbf{k}}{(2\pi)^3}\,
e^{i\mathbf{k}\cdot\mathbf{r}}\,
f_{\sphere}(\mathbf{k})\, \mathbf{A}(\mathbf{k}) ~~  ,   \\ 
\label{eq:thestartingpoint}
\mathcal{F}^{-1}\!\left[ f_{\sphere}\, \mathbf{B}     \right](\mathbf{r})
= \int \frac{d^3 \mathbf{k}}{(2\pi)^3}\,
e^{i\mathbf{k}\cdot\mathbf{r}}\,
f_{\sphere}(\mathbf{k})\, \mathbf{B}(\mathbf{k}) ~~  ,   
\end{align}
where $f_{\sphere}(\mathbf{k})$ is the function that represents the multipole expansion of flow around a sphere. When it holds that the force around the sphere is uniform, this function is given by \cite{meissneroszer2025exactresultsdissipationsteady,kim2005microhydrodynamics} 
\begin{align}  \label{eq:thesingularequation}
    f_{\sphere}\!\left(  k_\perp , k_\parallel    \right) = j_0 \left(a  \sqrt{k_\perp^2     + k_\parallel^2 }\right)  ~~ ,  
\end{align}
where $j_n ( s ) $ is the spherical Bessel function of the first kind and $a$ is the sphere radius.
\subsection{$\mathbf{B}$-term}
We will first work out $\mathcal{F}^{-1}\!\left[ f_{\sphere}\, \mathbf{B}     \right](\mathbf{r})$. Let us introduce cylindrical coordinates as
\begin{align}
k_x &= k_\perp \cos k_\phi, \qquad
k_y  = k_\perp \sin k_\phi, \qquad
k_z = k_\parallel , \qquad x = \rho \cos \phi, \qquad
y = \rho \sin \phi, \qquad
z = z .
\end{align}
\eqref{eq:thestartingpoint} can then be written as
\begin{align}
\mathcal{F}^{-1}\!\left[ f_{\sphere}\, \mathbf{B} \right](\mathbf{r})
&=   \frac{1  }{(2\pi)^3 }
\int_0^{\infty} \! \mathrm{d}k_\perp\, k_\perp
\int_0^{2\pi} \! \mathrm{d}k_\phi\,
e^{i k_\perp \rho \cos(k_\phi-\phi)} 
\int_{-\infty}^{\infty} \! \mathrm{d}k_\parallel\,
\frac{
f\!\left(  k_\perp , k_\parallel    \right)
k_\parallel^2
     \mathbf k      \mathbf k   
e^{i k_\parallel z}
}{   ( k_\parallel^2 + k_{\perp}^2  )^3   } .
\end{align}
For $z>0$ we close the $k_\parallel$ contour in the upper half-plane; the arc contribution vanishes and only the triple pole at $k_\parallel=i k_\perp$ contributes. Thus,
\begin{align}
\mathcal{F}^{-1}\!\left[ f_{\sphere}\, \mathbf{B} \right](\mathbf{r})
&=   \frac{i   }{(2\pi)^2 }
\int_0^{\infty} \! \mathrm{d}k_\perp\, k_\perp
\int_0^{2\pi} \! \mathrm{d}k_\phi\,
e^{i k_\perp \rho \cos(k_\phi-\phi)}   \frac{1}{2}   \left.
\left[  \frac{\partial^2 }{ \partial k_\parallel^2   }
\frac{
f\!\left(  k_\perp , k_\parallel    \right)
 k_\parallel^2  \mathbf k      \mathbf k  
 \,
e^{i k_\parallel z}
}{ (k_\parallel+i k_\perp  )^3 }
\right]
\right|_{k_\parallel=i k_\perp}.
\end{align}
which simplifies to
\begin{align}
\begin{split}
\mathcal{F}^{-1}\!\left[ f_{\sphere}\, \mathbf{B} \right](\mathbf{r})
&=   \frac{1  }{(2\pi)^2 }
\int_0^{\infty} \! \mathrm{d}k_\perp\, k_\perp
\int_0^{2\pi} \! \mathrm{d}k_\phi\,
e^{i k_\perp \rho \cos(k_\phi-\phi)}  e^{-  k_{\perp } z}\,
 \\  & \bigg[      \mathbf k_{\perp}      \mathbf k_{\perp}       M_3 (k_\perp , z )  +   ( \mathbf k_{\perp}     \bm \ell  +   \bm \ell    \mathbf k_{\perp}       )  \left(   i k_{\perp }    M_3 (k_\perp , z )  + 2   M_2 (k_\perp , z )  \right)  \\  & 
 +   (     \bm \ell     \bm \ell    \left(   - k_{\perp }^2    M_3 (k_\perp , z )  + 4   i k_{\perp }  M_2 (k_\perp , z )  + 2    M_1 (k_\perp , z )  \right) \bigg]   ~~,  
\end{split}  \label{eq:thingthing}
\end{align}
 Accounting for \eqref{eq:thesingularequation}, functions in \eqref{eq:thingthing} then turn into
\begin{subequations}
\begin{align}
        M_1 (k_\perp , z )  & =  \frac{1}{16 k_{\perp}}~~ , \\
    M_2 (k_\perp , z )  & = - \frac{i  \left(2 a^2 k_{\perp}^2-6 k_{\perp} z+3\right)}{96 k_{\perp}^2} ~~ ,  \\
    M_3 (k_\perp , z )  & =  - \frac{ \left(a^4 k_{\perp}^4-10 a^2 k_{\perp}^2 (k_{\perp} z-1)+15 \left(k_{\perp}^2 z^2-k_{\perp} z-1\right)\right)}{240 k_{\perp}^3}  ~~ . 
\end{align}
\end{subequations}

\subsection{$\mathbf{A}$-term} 
\label{eq:Bterm}
Let us now work out $\mathcal{F}^{-1}\!\left[ f_{\sphere}\, \mathbf{A}     \right](\mathbf{r})$. \eqref{eq:thestartingpoint2} can then be written as
\begin{align}
\mathcal{F}^{-1}\!\left[ f_{\sphere}\, \mathbf{A} \right](\mathbf{r})
&=   \frac{1  }{(2\pi)^3 }
\int_0^{\infty} \! \mathrm{d}k_\perp\, k_\perp
\int_0^{2\pi} \! \mathrm{d}k_\phi\,
e^{i k_\perp \rho \cos(k_\phi-\phi)} 
\int_{-\infty}^{\infty} \! \mathrm{d}k_\parallel\,
\frac{
f\!\left(  k_\perp , k_\parallel    \right)
     \mathbf k      \mathbf k   
e^{i k_\parallel z}
}{   ( k_\parallel^2 + k_{\perp}^2  )^2   } .
\end{align}
For $z>0$ we close the $k_\parallel$ contour in the upper half-plane; the arc contribution vanishes and only the double pole at $k_\parallel=i k_\perp$ contributes. Thus,
\begin{align}
\mathcal{F}^{-1}\!\left[ f_{\sphere}\, \mathbf{A} \right](\mathbf{r})
&=   \frac{i   }{(2\pi)^2 }
\int_0^{\infty} \! \mathrm{d}k_\perp\, k_\perp
\int_0^{2\pi} \! \mathrm{d}k_\phi\,
e^{i k_\perp \rho \cos(k_\phi-\phi)}    \left.
\left[  \frac{\partial }{ \partial k_\parallel  }
\frac{
f\!\left(  k_\perp , k_\parallel    \right)
   \mathbf k      \mathbf k  
 \,
e^{i k_\parallel z}
}{ (k_\parallel+i k_\perp  )^2 }
\right]
\right|_{k_\parallel=i k_\perp}.
\end{align}
which simplifies to
\begin{align}
\begin{split}
\mathcal{F}^{-1}\!\left[ f_{\sphere}\, \mathbf{A} \right](\mathbf{r})
&=   \frac{1  }{(2\pi)^2 }
\int_0^{\infty} \! \mathrm{d}k_\perp\, k_\perp
\int_0^{2\pi} \! \mathrm{d}k_\phi\,
e^{i k_\perp \rho \cos(k_\phi-\phi)}  e^{-  k_{\perp } z}\,
 \\  & \bigg[      \mathbf k_{\perp}      \mathbf k_{\perp}     N_2  (k_\perp , z )  +   ( \mathbf k_{\perp}     \bm \ell  +   \bm \ell    \mathbf k_{\perp}       )  \left(   i k_{\perp }    N_2  (k_\perp , z )  +   N_1  (k_\perp , z )  \right)  \\  & 
 +   (     \bm \ell     \bm \ell    \left(   - k_{\perp }^2    N_2  (k_\perp , z )  + 2   i k_{\perp }  N_1 (k_\perp , z )   \right) \bigg]   ~~.  
\end{split}  \label{eq:todo1}
\end{align}
Again taking \eqref{eq:thesingularequation}, we have
\begin{align}
        N_1 (k_\perp , z )   =  - \frac{i }{4 k_{\perp}^2} ~~ , ~~  N_2 (k_\perp , z )   =    - \frac{ \left(a^2 k_{\perp}^2-3 k_{\perp} z-3\right)}{12 k_{\perp}^3} ~~ . \end{align}
We integrating \eqref{eq:thingthing} and \eqref{eq:todo1} and using \eqref{eq:rewrite123}, we find
\begin{align}
   \mathbf{G}_{\sphere}^{ (\odd),\text{ANI}} (\mathbf{r})  = \mathcal{F}^{-1}\!\left[ f_{\sphere}\,    \tilde{\mathbf{G}}^{ (\odd),\text{ANI}}   \right](\mathbf{r})   ~~ . 
\end{align}
Combining with \eqref{eq:combinedthing}, we can obtain the odd viscous correction to the fluid profile
\begin{align}  
  \mathbf{u}^{(\odd)}  & =  -  \mathbf{ G}^{(\even) }  \cdot \, \mathbf{F}^{(\odd  )}_{\sphere }    - \left( \mathbf{ G}^{(\odd) ,\text{ISO}} +    \mathbf{ G}^{(\odd) ,\text{ANI}}  \right) \cdot \, \mathbf{F}^{(\even )}_{\sphere } ,   
\end{align}
where we note that $  \mathbf{F}^{(\even , \odd  )}_{\sphere}$ is the force on the spheroid induced by the fluid, hence the minus sign. 
To obtain $\mathbf{F}^{(o)}_{\sphere }$, we impose the boundary condition at first order in $\epsilon$, which implies
\begin{align}  
  \mathbf{u}^{(\odd)} (r , \theta ,  \phi ) \big|_{r =a}  & =  0 .    
\end{align}

\section{Singularity method for isotropic odd viscous flow around an oblate spheroid}
\label{eq:nonperturb}
We follow the same strategy as in App.~\ref{app:IFTsection}. Because for this computation we consider the isotropic case where $\gamma_\perp = \gamma_\parallel = \gamma$, we only need to obtain
\begin{align}
\label{eq:thestartingpoint23}
\mathcal{F}^{-1}\!\left[ f_{\obl}\, \mathbf{A}     \right](\mathbf{r})
= \int \frac{d^3 \mathbf{k}}{(2\pi)^3}\,
e^{i\mathbf{k}\cdot\mathbf{r}}\,
f_{\obl}(\mathbf{k})\, \mathbf{A}(\mathbf{k}) ~~  , 
\end{align}
where we have \cite{kim2005microhydrodynamics}
\begin{align}  \label{eq:thesingularequationspheroidal1}
    f_{\obl}\!\left(  k_\perp , k_\parallel    \right) = j_0 \left(  \sqrt{  a^2  k_\perp^2     + b^2 k_\parallel^2 }\right)  ~~ ,  ~~  a > b  ~~ . 
\end{align}
Like in App.~\ref{eq:Bterm}, we find 
\begin{align}
\begin{split}
\mathcal{F}^{-1}\!\left[ f_{\obl}\, \mathbf{A} \right](\mathbf{r})
&=   \frac{1  }{(2\pi)^2 }
\int_0^{\infty} \! \mathrm{d}k_\perp\, k_\perp
\int_0^{2\pi} \! \mathrm{d}k_\phi\,
e^{i k_\perp \rho \cos(k_\phi-\phi)}  e^{-  k_{\perp } z}\,
 \\  & \bigg[      \mathbf k_{\perp}      \mathbf k_{\perp}     O_2  (k_\perp , z )  +   ( \mathbf k_{\perp}     \bm \ell  +   \bm \ell    \mathbf k_{\perp}       )  \left(   i k_{\perp }    O_2  (k_\perp , z )  +   O_1  (k_\perp , z )  \right)  \\  & 
 +   (     \bm \ell     \bm \ell    \left(   - k_{\perp }^2    O_2  (k_\perp , z )  + 2   i k_{\perp }  O_1 (k_\perp , z )   \right) \bigg]   ~~.  
\end{split}  \label{eq:todo}
\end{align}
With \eqref{eq:thesingularequationspheroidal1}, we find
\begin{align}
        O_1 (k_\perp , z )    &  =   -\frac{i \, \text{sinc}\left(k_{\perp } \alpha\right)}{4 k_{\perp }^2} ~~ ,   \\ 
        O_2 (k_\perp , z )    &  =   \frac{\left(a^2 (k_{\perp } z+1)-b^2 (k_{\perp } z+2)\right) \sin \left(k_{\perp } \alpha\right)+b^2 k_{\perp } \alpha  \cos \left(k_{\perp } \alpha\right)}{4 k_{\perp }^4 \alpha^3 } ~~ ,  \end{align}
        where $\alpha = \sqrt{a^2 - b^2 }$. 
        Integrating \eqref{eq:thingthing} and \eqref{eq:todo1} and using \eqref{eq:isoiso}, we find
\begin{align}
   \mathbf{G}_{\obl}^{ (\odd),\text{ISO}} (\mathbf{r})  = \mathcal{F}^{-1}\!\left[ f_{\obl}\,    \tilde{\mathbf{G}}^{ (\odd),\text{ISO}}   \right](\mathbf{r})   ~~ . 
\end{align}
To verify the result from the integration of \eqref{eq:todo}, we note that the reciprocal Stokeslet solution of \eqref{eq:stokeslet} can be written as   
\begin{equation}  \label{eq:rewrite12}
\tilde{\mathbf{G}}^{(\even)}   (\mathbf{k} )=\frac{1}{\eta_\sh }   \left[   \mathbf{1} \text{Tr} \left( \mathbf{A} \right)    - \mathbf{A}  \right]    ~~ , 
\end{equation}
which means that $\mathcal{F}^{-1}\!\left[ f_{\obl}\, \mathbf{A} \right](\mathbf{r})$ can be directly used to obtain the even viscous flow $ \mathbf{u}^{(0) } $ as well. For this velocity, we require that \cite{everts2023dissipative}
\begin{align} \label{eq:evertsmasterequation}
     \mathbf{u}^{(0) }  ( \rho  , \phi , z   ) \bigg|_{\rho = a \cos(\theta ) , z = b  \sin(\theta )  } =  -     \mathbf{G}_{\obl}^{ (\odd)}  ( \rho  , \phi , z   ) \bigg|_{\rho = a \cos(\theta ) , z = b  \sin(\theta )  }   \cdot \mathbf{F}^{(\even )}_{\obl} = \mathbf{U} ~~   . 
\end{align}
Solving \eqref{eq:evertsmasterequation} yields 
\begin{align}
    \mathbf{F}^{(\even )}_{\obl}   = -   \left[ \zeta^{(\even) , \perp }_{\obl}  (  \mathbf{1} - \bm \ell \bm \ell )   +   \zeta^{(\even) , \parallel  }_{\obl}  \bm \ell \bm \ell \right] \cdot \mathbf{U}  ~~ . 
\end{align}
with 
\begin{subequations}
    \begin{align}
  \zeta^{(\even) , \perp }_{\obl}  &  =   \frac{16 \pi  \eta_\sh   (a^2-b^2)^{3/2}  }{  - b \sqrt{a^2-b^2} + 3 a^2 \tan ^{-1}\left(\frac{\sqrt{a^2-b^2}}{b}\right)- 2 b^2 \tan ^{-1}\left(\frac{\sqrt{a^2-b^2}}{b}\right)} ~~ ,  \\ 
   \zeta^{(\even) , \parallel  }_{\obl} &  =  \frac{8 \pi  \eta_\sh \left(a^2-b^2\right)^{3/2}}{b \sqrt{a^2-b^2}+\left(a^2-2 b^2\right) \cot ^{-1}\left(\frac{b}{\sqrt{a^2-b^2}}\right)} ~~ . 
    \end{align}
\end{subequations}
This result is identical to what is found in \cite{kim2005microhydrodynamics}. The odd correction to the fluid profile is given by 
\begin{align}
   \mathbf{u}^{(\odd)}  =   -    \mathbf{G}_{\obl}^{ (\even) }   \cdot \mathbf{F}^{(\odd )}_{\obl}    -       \mathbf{G}_{\obl}^{ (\odd),\text{ISO}}    \cdot \mathbf{F}^{(\even )}_{\obl}   ~~ . 
\end{align}
We then require that
\begin{align}
   \mathbf{u}^{(\odd)}   ( \rho  , \phi , z   ) \big|_{\rho = a \cos(\theta ) , z = b  \sin(\theta )  }   =0 
\end{align}
Solving, we find 
\begin{align}
    \mathbf{F}^{(\odd  )}_{\obl}   =  -    \zeta^{(\odd )}_{\obl} \bm \varepsilon  \cdot \mathbf{U}      ~~ ,  
\end{align}
with 
\begin{align}
 \zeta^{(\odd )}_{\obl}  = 32 \pi \gamma   \frac{ (a^2-b^2)^{3/2} \left(a^2 \cot ^{-1}\left(\frac{b}{\sqrt{a^2-b^2}}\right)   - b \sqrt{a^2-b^2} \right)}{\left(\left(2 b^2-3 a^2\right) \cot ^{-1}\left(\frac{b}{\sqrt{a^2-b^2}}\right)+b \sqrt{a^2-b^2}\right)^2} ~~ . 
\end{align}
which is \eqref{eq:bigresult}.

\end{document}